\documentclass{aa}  

\usepackage{hyperref}
\usepackage{graphicx}
\usepackage{txfonts}
\usepackage[normalem]{ulem}
\usepackage{xspace} 

\newcommand{\HIMF}{H\textsc{i}MF\xspace}
\newcommand{\HI}{H\,\textsc{i}\xspace}
\newcommand{\Msol}{$M_{\odot}$\xspace}
\newcommand{\Mstar}{$M_\mathrm{\star}$\xspace}
\newcommand{\MHI}{$M_\mathrm{HI}$\xspace}

\newcommand{\kms}{km\,s$^{-1}$\xspace}
\newcommand{\Mvir}{$M_\mathrm{vir}$\xspace}
\newcommand{\Rvir}{$R_\mathrm{vir}$\xspace}
\newcommand{\vsys}{$v_{\rm sys}$\xspace}

\newcommand{\degree}{$^\circ$\xspace}
\newcommand{\SoFiA}{\texttt{SoFiA}\xspace}

\newcommand{\cm}{cm$^{-2}$\xspace}
\newcommand{\Tc}{$T_{c}$\xspace}

\begin{document}

   \title{The MeerKAT Fornax Survey}
    \subtitle{VI. The collapse of the galaxy \HI Mass Function in Fornax}

   \titlerunning{Fornax Cluster \HIMF}

   \author{D.~Kleiner\inst{1, 2}, P.~Serra\inst{2}, A.~Loni\inst{3}, S.~H.~A.~Rajohnson\inst{2, 4}, F.~M.~Maccagni\inst{2}, W.~J~.G.~de Blok\inst{1, 4, 5}, P.~Kamphuis\inst{6}, R.~C.~Kraan-Korteweg\inst{4} \and M.~A.~W.~Verheijen\inst{5}}
   \authorrunning{Kleiner et al.}

  \institute{Netherlands Institute for Radio Astronomy (ASTRON), Oude Hoogeveensedijk 4, 7991 PD Dwingeloo, the Netherlands\\
              \email{kleiner@astron.nl}
              \and
              INAF - Osservatorio Astronomico di Cagliari, Via della Scienza 5, I-09047 Selargius (CA), Italy
              \and
              INAF - Osservatorio Astronomico di Capodimonte, Salita Moiariello, 16, 80131 Napoli (NA), Italy
              \and
              Department of Astronomy, University of Cape Town, Private Bag X3, Rondebosch 7701, South Africa
              \and
              Kapteyn Astronomical Institute, University of Groningen, PO Box 800, NL-9700 AV Groningen, the Netherlands
              \and
              Ruhr University Bochum, Faculty of Physics and Astronomy, Astronomical Institute (AIRUB), 44780 Bochum, Germany
}

   \date{Received 4 August 2025 / Accepted 20 November 2025}

  \abstract{We present the deepest \HI mass Function (\HIMF) ever measured, outside the Local Group. The observations are part of the MeerKAT Fornax Survey and cover a 4 $\times$ 4 deg$^{2}$ field, corresponding to $\sim$ \Rvir. The 3$\sigma$ detection limit is log(\MHI/\Msol) = 5.7 for a 50\,\kms-wide point source. We detect \HI in 35 galaxies and 44 clouds with no optical counterparts. Using deep optical images from the Fornax Deep Survey, we measure the 5$\sigma$ optical flux limit of the \HI clouds and show that they are a distinct population, separated by a four magnitude gap from the faintest \HI-detected galaxies. Three quarters (33 out of 44) of the clouds are associated with the two galaxies with the most \HI in the cluster -- NGC\,1365 and NGC\,1427A, although the clouds contribute a negligible amount to the total \MHI budget. By performing a signal-to-noise analysis and computing the Rauzy statistic on the \HI detections, we demonstrate that our catalogue is complete down log(\MHI/\Msol) = 6, and we are therefore readily able to probe the \HIMF down to this level. We find an abrupt drop of the number density of \HI-detected galaxies at log(\MHI/\Msol) = 7, signifying a clear absence of galaxies between 6 $<$ log(\MHI/\Msol) $\leq$ 7. We use the modified maximum likelihood method to fit a Schechter function down to log(\MHI/\Msol) $\geq$ 7, the range where the \HIMF follows a power-law. The measured low-mass slope is $\alpha$ = $-1.31$ $\pm$ 0.13, with a characteristic knee mass of log(M$_{*}$/\Msol) = 10.52 $\pm$ 1.89. The low-mass slope matches the slope in the field, while the knee is defined by a single galaxy and is unconstrained. Below log(\MHI/\Msol) = 7, there is a sharp departure from a Schechter function, and we report the first robust measurement of the collapse of a \HIMF. For the \HIMF below log(\MHI/\Msol) = 7 to follow a power-law, tens of galaxies are needed -- a factor $\sim$ six higher than what is observed. The collapse of the Fornax \HIMF is likely due to the rapid removal of \HI from low-mass galaxies.}

   \keywords{galaxies: evolution -- galaxies: clusters: individual: Fornax -- galaxies: ISM -- galaxies: luminosity function, mass function -- radio lines: galaxies}

   \maketitle

\section{Introduction}
Atomic hydrogen is the building block for all baryonic matter in the Universe. It is prevalent at all scales, from individual clouds that collapse and form stars, to the dominant material of the Cosmic Web. In the context of galaxy evolution, neutral atomic hydrogen (\HI) serves as the primary component of the interstellar medium (ISM). As the lightest and most weakly bound element of the cold ISM, \HI is the most sensitive tracer of internal and external physical processes that dictate the life cycle of galaxies. 

\HI is particularly sensitive to environmental interactions. There is an abundance of evidence from observations and simulations showing that \HI is affected by hydrodynamical and tidal forces before other components, such as the molecular gas and stellar population of galaxies \citep[e.g.][]{Chung2009b, Morasco2016, Catinella2018, deBlok2018, Kleiner2021, Kleiner2023, Rohr2023}. By observing \HI across a range of environments, it has been established that the removal of the \HI reservoir prevents future star formation and ultimately leads to quenching of a galaxy \citep[e.g. see][for extensive reviews]{Boselli2006, Boselli2014c, Cortese2021, Boselli2022}. Besides studying the \HI properties of individual galaxies, we can gain fundamental insights into the effect of environment by quantifying the properties of the galaxy population within a given volume.  

One method is to estimate the mass or luminosity function, which quantifies the number density of galaxies as a function of mass or luminosity, and serves as a fundamental observable for constraining the Lambda Cold Dark Matter ($\Lambda$CDM) model of the Universe. These functions act as statistical benchmarks in simulations, allowing comparisons of different baryonic physics models (such as star formation and feedback) with observations that provide key constraints on galaxy formation and evolution. \citep[e.g. see][]{Jenkins2001, Benson2003, Baldry2004, Loveday2012, Joop2015}. Luminosity and mass functions have been traditionally fitted by the Schechter function \citep{Schechter1976}. When applied to the number density of \HI-detected galaxies per unit volume, the Schechter function takes the form:

\begin{equation}
\phi(M_{\rm HI}) = \frac{dn}{d\rm{log}(M_{\rm HI})} = \rm{ln(10)}\phi_*\left(\frac{M_{\rm HI}}{M_*}\right)^{\alpha+1} \rm{exp}\left(-\frac{M_{\rm HI}}{M_*}\right)
\end{equation}

where the observed number density per \HI mass bin $\phi(M_{\rm HI})$, depends on the normalisation $\phi_*$, the low-mass slope $\alpha$ and the characteristic `knee' mass M$_*$. 

Unsurprisingly, the \HI Mass Function (\HIMF) is sensitive to environmental effects \citep[e.g.][]{Schneider1997, Zwaan1997, Zwaan2005, Springob2005, Moorman2014, Jones2016, Said2019}, making it a powerful tool for studying galaxy interactions across different environments. At z $\sim$ 0, the field \HIMF has been measured using the \HI Parkes All Sky Survey \citep[HIPASS;][]{Barnes2001, Meyer2004}, the Arecibo Legacy Fast Arecibo L-band Feed Array survey \citep[ALFALFA;][]{Giovanelli2005, Haynes2011}, and more recently the Five-hundred-meter Aperture Spherical radio Telescope (FAST) All Sky \HI survey \citep[FASHI;][]{Zhang2024} and the Looking at the Distant Universe with the MeerKAT Array \citep[LADUMA;][]{Blyth2016} survey. In general, these surveys show good agreement -- consistently determining a field \HIMF slope of $\alpha$ = $-1.3$ \citep{Zwaan2003, Zwaan2005, Jones2018, Ma2025, Kazemi-Moridani2025}. However, it is worth noting that a modest difference in the ALFALFA slope ($\Delta\alpha$ = 0.14) was reported between Northern and Southern hemispheres, likely due to cosmic variance \citep{Jones2018}, and the measured LADUMA slope ($\alpha$ = $-1.2$) is slightly flatter although consistent within the uncertainty. In the group environment, a range of slopes has been reported. Most studies find a slope between $-1.3$ and $-1.0$ \citep[e.g.][]{Verheijen2001, Kovac2007, Freeland2009, Kilborn2009, Pisano2011, Westmeier2017, Jones2020, Busekool2021, Yu2025} and hint at a tentative trend towards a flatter $\alpha$ than in the field. Others report the same slope ($\alpha$ = $-1.3$) as the field \citep{ Martin2010, Jones2016}, while steeper values ($\alpha$ $<$ $-1.4$) have also been observed \citep{Stierwalt2009, Davies2011}. This broad variation in $\alpha$ within groups is likely related to the timescale of \HI removal, which depends on the density of the intragroup medium, galaxy number density, and the typical speed of galaxies relative to one another and to the medium. While $\alpha$ describes the \HIMF faintwards of the knee, the \HIMF in the low-mass regime i.e. log(\MHI/\Msol) $\leq$ 7 is largely unconstrained. Almost all previous studies have been limited to log(\MHI/\Msol) $\gtrsim$ 7 and the few that probe below log(\MHI/\Msol) = 7 suffer from large uncertainties due to completeness corrections. The primary limitation has been the lack of simultaneous sensitivity and resolution needed to construct a complete, volume-limited \HI catalogue below log(\MHI/\Msol) = 7. In this mass range, galaxies can rapidly lose their \HI and the shape of the \HIMF has not yet been precisely measured.
 
The Fornax cluster is a small, low-mass cluster (\Mvir = 5 $\times$ 10$^{13}$\,\Msol) located in the Southern sky \citep{Drinkwater2001}. In comparison to Virgo \citep[e.g.][]{Jordan2007} which is the most nearby cluster, the Fornax galaxy number density is twice as high, although the intracluster medium (ICM) is half as dense \citep{McCall2024, Reiprich2025}. As Fornax is only 20\,Mpc away \citep{Blakeslee2009}, it is an excellent target for detailed studies of galaxy transformations and evolution. The population of Fornax is dominated by red, low-mass quiescent galaxies. The cluster clearly is not viralized yet as demonstrated by the highly asymmetric spatial distribution of massive early-type galaxies, the sloshing of the ICM \citep{Machacek2005, Scharf2005, Su2017, Sheardown2018}, and in-falling groups that include massive star-forming late type galaxies \citep{Drinkwater2001, Raj2019, Kleiner2021, Loni2021}. Previous studies \citep[such as][]{Drinkwater2001, Schroder2001, Waugh2002, Iodice2016, Iodice2017, Venhola2018, Lee-Waddell2018, Raj2019, Zabel2019, Spavone2020, Spavone2022, Loni2021} have reported the effects of hydrodynamical and tidal forces operating in the cluster. More recently, multi-wavelength studies using the powerful combination of deep optical imaging and resolved, low column density \HI observations are revealing new insights. The Fornax Deep Survey \citep[FDS;][]{Iodice2016, Venhola2018, Peletier2020} observed the Fornax cluster using the VLT Survey Telescope (VST) out to the virial radius (\Rvir) and beyond, covering the infalling NGC\,1316 group in the Sloan Digital Sky Survey (SDSS) \textit{ugriz} bands. At 1\,arcsec resolution, the FDS is sensitive down to $\mu_{g}$ = 28.4\,mag\,arcsec$^{-2}$, and has proven to be excellent at detecting faint dwarfs and resolving low surface brightness features \citep[e.g.][]{Iodice2016, Venhola2018, Venhola2019, Su2021}. The MeerKAT telescope \citep{Camilo2018} provides the required sensitivity and resolution, along with the ability to detect and identify sources below log(\MHI/\Msol) = 7 at the distance of Fornax. Taking advantage of this new and unexplored parameter space, the MeerKAT Fornax Survey \citep[MFS;][]{Serra2023} has completed radio continuum and 21\,cm spectral line observations with a similar footprint to the FDS. The MFS is able to detect below log(\MHI/\Msol) = 6 for an unresolved source with a 50\,kms line-width in the Fornax cluster and offers the deepest, most resolved \HI images of the Fornax cluster to date \citep{Serra2023}. Several recent studies combine the FDS and MFS data to gain better insights into the details of \HI stripping. For example, \citet{Serra2023} showed six galaxies with disturbed stellar bodies that host radially oriented, long \HI tails with no stellar component. These features were produced by tidal forces initially displacing \HI from the stellar body, followed by ram pressure shaping the weaker bound \HI. By measuring the star-formation history of NGC\,1436, a lenticular galaxy in the making, \citet{Loni2023} showed that the \HI was quickly stripped from the outer disk, causing rapid quenching, and is now confined to the inner star-forming disk, deep within the stellar body. The work of \citet{Serra2024} derived a detailed kinematic model of NGC\,1427A, the galaxy with the longest \HI tail \citep[first reported in][]{Lee-Waddell2018} in the cluster, and found strong evidence of a two-step interaction where tidal and hydrodynamical interactions are both required to explain the complex kinematics. Lastly, \citet{Kleiner2023} conducted a detailed analysis on the individual, low-mass \HI-detected galaxies in the Fornax cluster. They used a toy model to match the observed number of late type dwarf galaxies with \HI and showed that dwarf galaxies are stripped within a few hundred Myr. Due to their low \MHI and shallow potential, dwarfs below log(\MHI/\Msol) = 7 experience instantaneous \HI stripping in their first encounter, whether it from tidal, hydrodynamical or both forces in the cluster.  

In this paper, we make use of all the MFS \HI detections within $\sim$ \Rvir and present the first \HIMF of the Fornax cluster. We detect galaxies with \HI and \HI clouds in the ICM down to log(\MHI/\Msol) = 5.7. Our main goal is to measure $\alpha$ and unambiguously determine the shape of the low-mass end of the \HIMF in the Fornax cluster. The \HIMF presented in this work is the deepest ever measured, outside the Local Group.  

This paper is arranged as follows: Section \ref{sec:catalogue} presents the \HI catalogue. We discuss the differences between \HI-detected galaxies and free-floating clouds, and explore the completeness of the optical and \HI catalogues. In Section \ref{sec:HIMF} we present the \HIMF of the Fornax cluster, and discuss what type of galaxies make up the \HIMF. Section \ref{sec:comparison} is where we compare our \HIMF to others and discuss the implications of the new findings. Our results are summarised in Section \ref{sec:conclusion}. Throughout this paper, we assume that Fornax cluster galaxies are at a distance of 20\,Mpc \citep{Blakeslee2009}.

\section{MeerKAT Fornax Survey \HI catalogue}
\label{sec:catalogue}
The $\sim$ 4 $\times$ 4 deg$^{2}$ field in this work covers the virial radius (\Rvir $\sim$ 2\degree, 700\,kpc) of the Fornax cluster, while excluding the massive infalling Fornax A group. We use the 41\arcsec-resolution cube \citep[described in][]{Serra2023}, that was created using a robust value of 0.5 and a uv-taper of 30\arcsec, resulting in a restoring beam of 43\arcsec $\times$ 40\arcsec, corresponding to $\sim$ 4\,kpc at 20\,Mpc, the distance of Fornax. This cube offers the highest point source sensitivity, yielding the lowest 3$\sigma$ log($M_\mathrm{HI,lim}$/\Msol) = 5.7 for a source with a velocity width of 50\,\kms. We use \SoFiA \citep{Serra2015a, Westmeier2021} to detect and parametrise the \HI sources. Gaussian spatial kernels with FWHM values up to 6 pixels and boxcar spectral kernels up to 61 channels wide were applied using a 4$\sigma$ detection threshold. We visually inspected both the cube and individual detections to de-blend and merge relevant sources, producing the final mask. This mask was then used as the clean mask, for generating moment maps and for the  parametrisation of the \HI sources.

\begin{figure}
    \centering
    \includegraphics[width=1\linewidth]{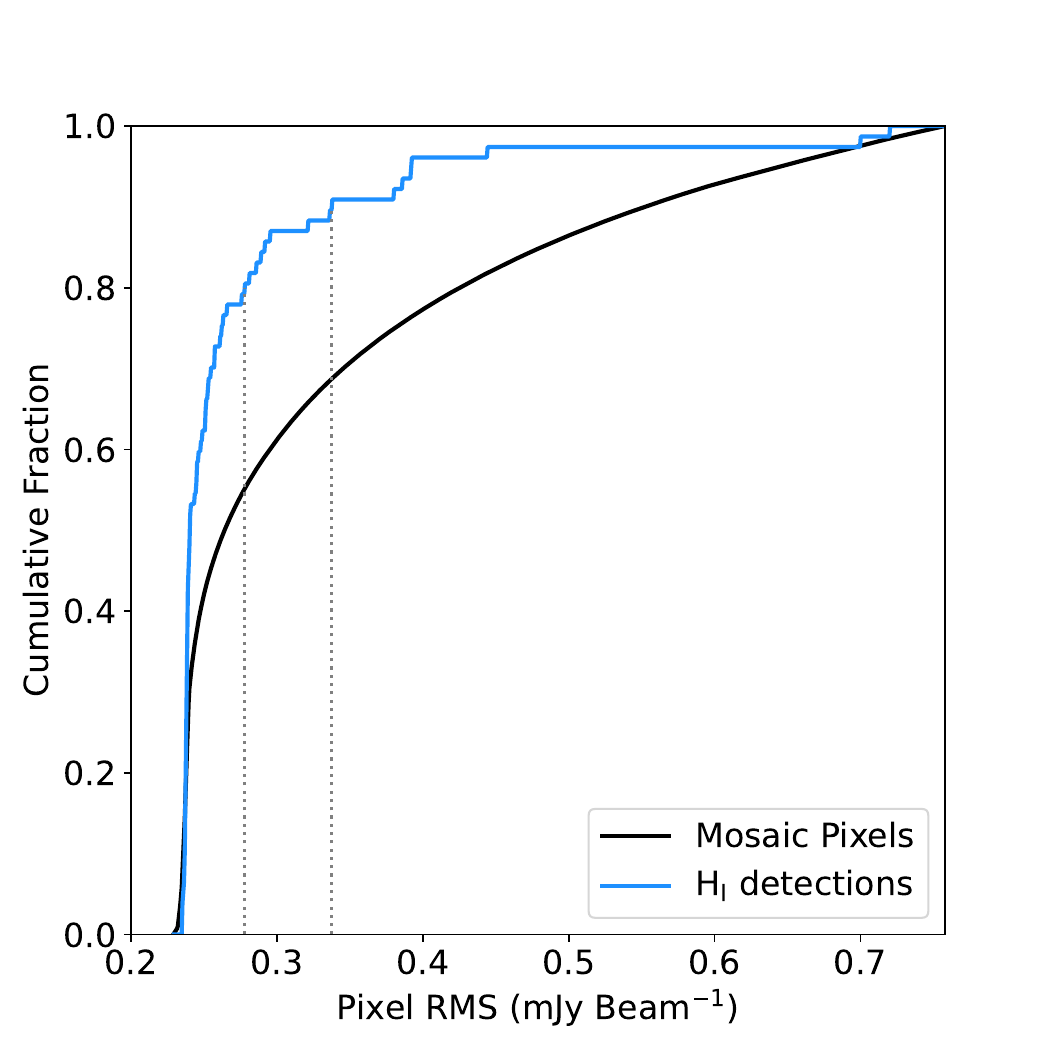}
    \caption{Cumulative histogram of the RMS values. The black and blue histograms show the cumulative distribution of the median pixel value measured along the line-of-sight in the mosaic, and the RMS at the location of each \HI-detection. The grey dotted lines at 0.28 and 0.34 \,mJy\,beam$^{-1}$ show the 80$^{\text{th}}$ and 90$^{\text{th}}$ percentiles of the cumulative distribution of \HI detections, respectively. The overwhelming majority of \HI detections detected near the minimum RMS of the mosaic, with only a small variation.}
    \label{fig:rms_pix}
\end{figure}

Fig.\,\ref{fig:rms_pix} shows the cumulative distribution of the median value of the root mean square (RMS) noise for every pixel in the mosaic, measured along the line-of-sight. The cumulative distribution of the RMS measured at the location of each \HI detection is also shown (blue line), along with the 80$^{\text{th}}$ and 90$^{\text{th}}$ percentiles (grey dotted lines) of the cumulative distribution of \HI detections. The central RMS of the mosaic is 0.24\,mJy\,Beam$^{-1}$, and the 80$^{\text{th}}$ and 90$^{\text{th}}$ percentiles of the detections intersect at 0.28 and 0.34\,mJy\,Beam$^{-1}$, respectively. This shows that 80\% (and 90\%) of the detections reside where the noise is higher than at the centre of the mosaic by only 20\% (and 50\%). The low variation in the local RMS at the positions of the \HI detections reflects that the vast majority are located far from the mosaic edges.

\subsection{\HI-detected galaxies and clouds}
\label{sec:galclouds}
There are 413 galaxies catalogued by the FDS \citep{Iodice2016, Venhola2018, Peletier2020, Su2021} within the area of the MFS mosaic used in this paper. A total of 35 galaxies are detected in \HI, and we present their \MHI, systemic velocity ($v_{\rm sys}$), velocity width at 20\% of the profile peak flux ($w_{20}$) and absolute $r$-band magnitude in Table \ref{tab:gal_detections}\footnote{ESO\,358-G060 has been excluded from the optical and \HI catalogues, as it has been deemed to be a foreground galaxy based on its position on the baryonic Tully-Fisher relation \citep{Kamphuis2025}}. Using a distance of $D$ = 20\,Mpc for all sources, the \HI masses were computed using equation 49 from \citet{Meyer2017}. Uncertainties are estimated by adding the statistical uncertainty and the 10\% flux-scale calibration uncertainty, in quadrature. 

In this work, we aim to measure the \HIMF of galaxies, making it essential to distinguish galaxies from free-floating \HI clouds in the ICM. A \HI cloud is identified when there is no optical counterpart in the FDS catalogue, nor any optical emission is visible in the single-band FDS images after excluding obvious background sources through visual inspection of an optical ($gri$) colour image. We identify 44 \HI clouds with no optical counterpart -- an unprecedented number within a small volume. The position, \MHI, $v_{\rm sys}$, $w_{20}$, name of nearest galaxy in projection with a measured spectroscopic (optical or \HI) redshift, and the systemic velocity difference ($\lvert\Delta v_{\rm sys}\rvert$) between the cloud and nearest galaxy are listed in Table\,\ref{tab:cloud_detections}. At this resolution, all the clouds are smaller than $\sim$ three beams, with the majority being of the order of a beam, and thus spatially unresolved. The typical $w_{20}$ line-width of the clouds is approximately 35\,\kms(Table~\ref{tab:cloud_detections}), comparable to that of some low-mass \HI-bearing dwarf galaxies. For each cloud, we extract a 4 $\times$ 4 arcmin$^{2}$ FDS $r$-band cutout, mask bright emission, and estimate the noise by fitting a Gaussian to the distribution of pixels below the mode. We then place a 5$\sigma$ upper limit on the $r$-band optical flux within a 20\arcsec\xspace aperture (corresponding to 2\,kpc). 

\begin{figure}
    \centering
    \includegraphics[width=1\linewidth]{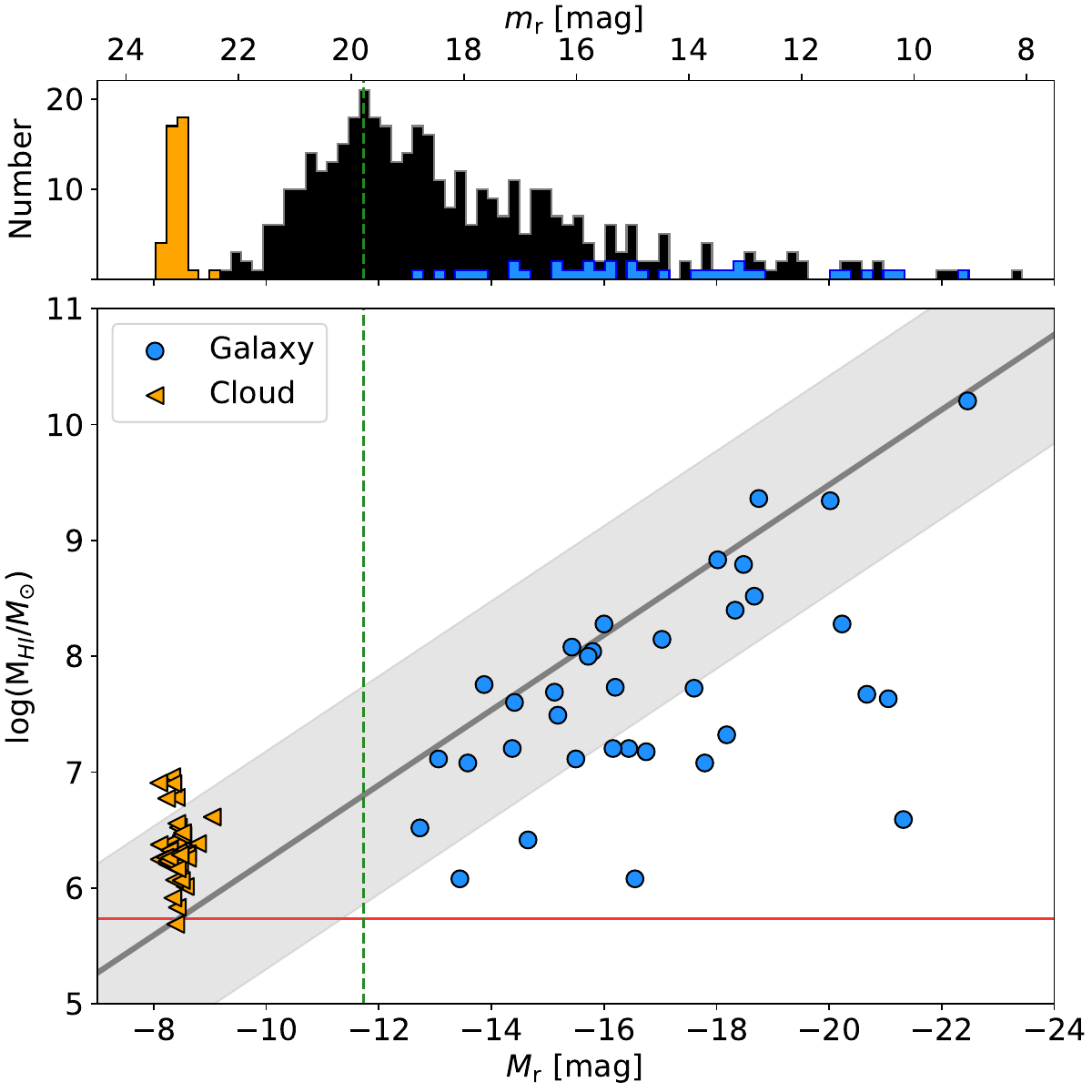}
    \caption{$M_{r}$-\MHI plane. The main figure shows galaxies detected in \HI as filled blue circles, and the \HI cloud 5$\sigma$ optical flux upper limit are shown as orange left pointing arrows. The grey line and shaded region is the mean and scatter of the scaling relation of CVn galaxies \citep{Kovac2009}, the red solid line represented the lowest \MHI detection limit of the mosaic, and the dashed green line at $M_{r}$ = $-11.7$ denotes the completeness limit of the FDS catalogue. The top panel shows absolute and apparent $r$-band histogram of all cluster galaxies (black), \HI-detected galaxies (blue), and the optical flux 5$\sigma$  upper limits of the \HI clouds. The \HI-detected galaxies and clouds are clearly two distinct populations.}
    \label{fig:Mr_MHI}
\end{figure}

In Fig.\,\ref{fig:Mr_MHI}, we show the \HI-detected galaxies and free-floating \HI clouds on the $M_{r}$-\MHI plane, relative to the $M_{r}$-\MHI relation and its associated scatter. This relation is defined by typical star-forming, gas-rich field galaxies (6.5 $<$ log(\MHI/\Msol) $\leq$ 9.8) in the Canes Venatici (CVn) volume \citep{Kovac2009, Kleiner2023}. The top panel of Fig.\,\ref{fig:Mr_MHI} presents the absolute (bottom axis) and apparent (top axis) $r$-band magnitude histograms of these \HI detections, compared to all catalogued FDS galaxies. From the FDS catalogue (black histogram in Fig.\,\ref{fig:Mr_MHI}), we determine the completeness limit to be $M_{r} = -11.7$\,mag (green dashed line), although 121 galaxies are catalogued up to two magnitudes fainter. All \HI-detected galaxies have log(\MHI/\Msol) $\gtrsim$ 6 and $M_{r} \lesssim -13$, making them more than a magnitude brighter than the $r$-band completeness limit. Most \HI clouds fall within 6 $<$ log(\MHI/\Msol) $\leq$ 7, a range where many galaxies are detected. However, there is a $\sim$ four magnitude gap between the 5$\sigma$ upper limits on the optical flux of the \HI clouds and the faint end of the \HI-rich galaxy population. In this parameter space, the clouds appear as a distinct population of optically ultra-faint, \HI-rich objects rather than a simple continuation of the galaxy population toward lower luminosities.

\begin{figure*}
    \centering
    \includegraphics[width=1\linewidth]{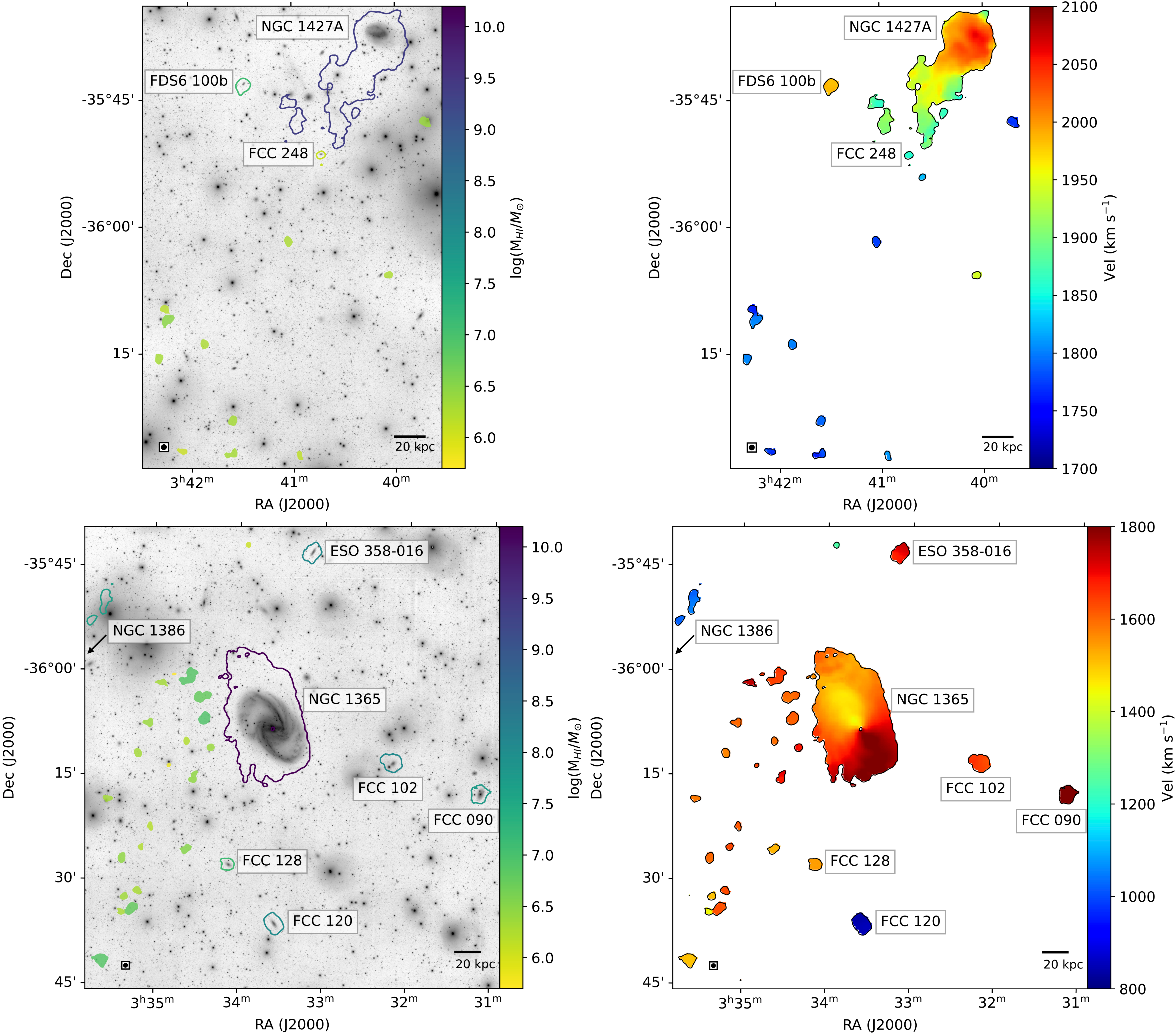}
    \caption{Optical, \HI emission and velocity field of NGC\,1427A (top) and NGC\,1365 (bottom). The left shows the FDS $g$-band image, with 3$\sigma$ $N_\mathrm{HI,25\,km\,s^{-1}}$ = 2.7 $\times$ 10$^{18}$ \cm contour overlaid. The \HI-detections are coloured according to their integrated \MHI, where galaxies are labelled and are shown with open contours, and \HI clouds are filled. On the right are the velocity fields of the respective fields. In all images, the 43\arcsec $\times$ 40\arcsec beam is shown in the bottom left and a 20\,kpc scale bar on the bottom right. With respect to NGC\,1427A and NGC\,1365, the \HI clouds have similar velocities, and show a continuous gradient.}
    \label{fig:MFS_overlay}
\end{figure*}

An important consideration is whether the \HI clouds should be included in the \MHI of any galaxy. In Fig.\,\ref{fig:MFS_overlay}, we show the fields of the two \HI-richest galaxies in the cluster: NGC\,1427A and NGC\,1365. These fields contain both high- and low-\MHI galaxies, as well as three quarters of the \HI clouds (33 out of 44). The clouds near NGC\,1427A extend up to $\sim$ 260\,kpc in projection from the optical centre of the galaxy, where they exhibit a continuous velocity gradient aligned with the galaxy’s long \HI tail, reinforcing the idea that they originate from its ongoing environmental \HI stripping \citep[see][for a detailed analysis]{Serra2024}. Although the origin of the clouds near NGC\,1365 has not yet been established, it is possible that similar cluster mechanisms are stripping \HI from this high-\MHI galaxy -- generating a $\sim$ 200\,kpc long stream of \HI clouds in the ICM. Given the close spatial and velocity ($\lesssim$ 200\,\kms of \vsys) proximity of the clouds to the two massive, \HI-rich galaxies, it is reasonable to associate these clouds with NGC\,1427A and NGC\,1365. These two galaxies have respective \HI masses of log(\MHI/\Msol) = 9.4 and 10.2 $\pm$ 0.1. The combined \MHI of their associated clouds is log(\MHI/\Msol) = 7.4 for NGC\,1427A and 7.8 for NGC\,1365 -- a negligible fraction compared to their total \MHI, well within their uncertainty. Consequently, the inclusion of these clouds in the total \MHI budget of these galaxies does not affect the Fornax \HIMF. For the remaining nine \HI clouds (Table\,\ref{tab:cloud_detections}), if we attributed them all to the nearest galaxy, our results  would not change. However, as their origins are unclear, we do not attempt to attribute them to any galaxy.

\subsection{\HI catalogue completeness}
\label{sec:completeness}
\begin{figure}
    \centering
    \includegraphics[width=1\linewidth]{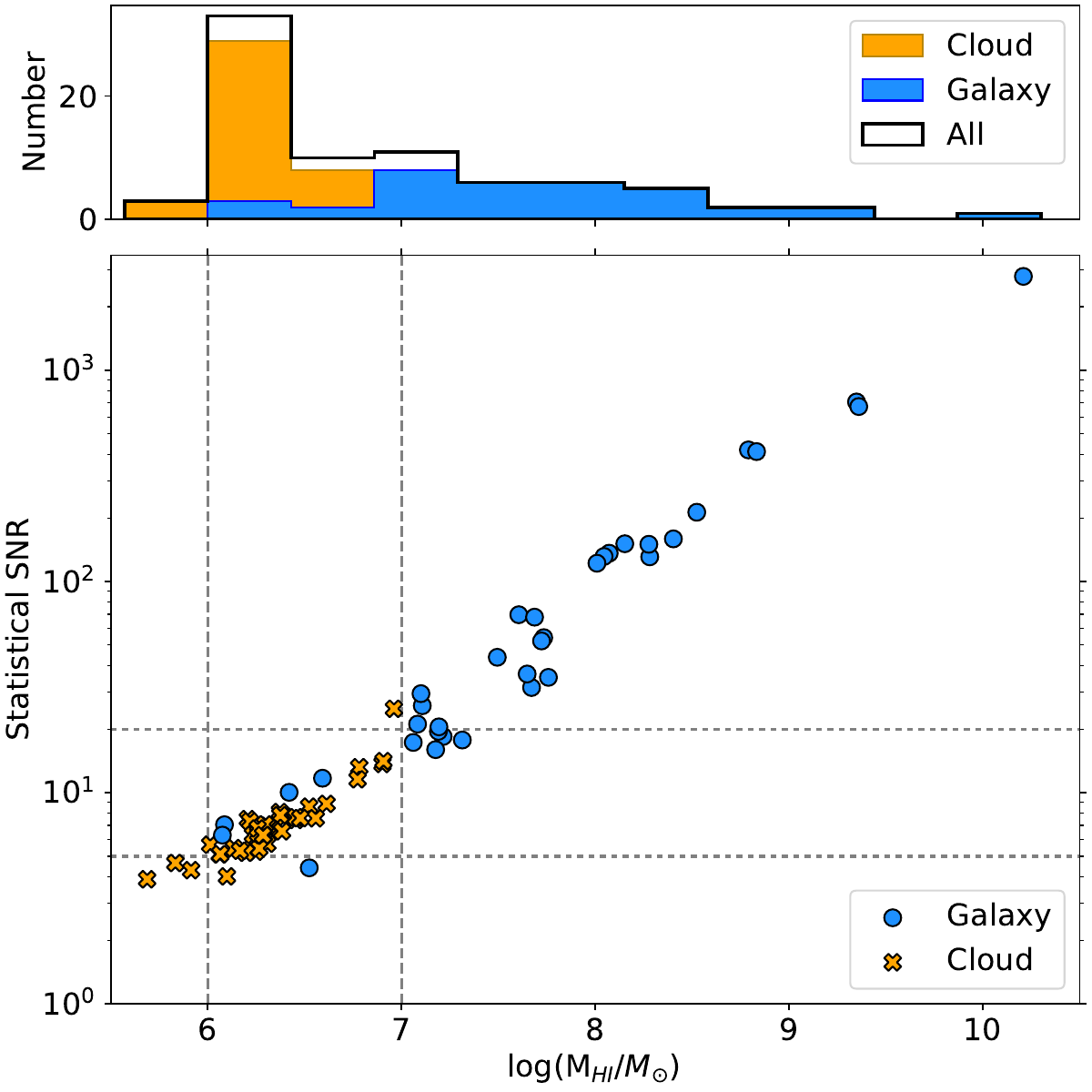}
    \caption{The \MHI and statistical SNR. In the main panel, the blue points and orange crosses show the \HI-detected galaxies and clouds respectively. Dashed lines have been placed at log(\MHI\Msol) = 6 and 7, and SNR = 5 and 20. The top panel shows the number of detections for all \HI-sources (black) and the galaxies and clouds using the same colours as the main panel. The top panel shows a severe truncation of \HI-detected galaxies at log(\MHI/\Msol) = 7, where the typical statistical SNR is 20.}
    \label{fig:HI_SNR}
\end{figure}

Understanding the completeness of the \HI catalogue is a crucial factor when probing the low-mass end of the \HIMF. We show the integrated signal-to-noise ratio (SNR) for all \HI-detections in Fig.\,\ref{fig:HI_SNR}, which was calculated by \SoFiA as the ratio between the total flux and the corresponding noise calculated within the 3D detection mask \citep[see Eq. 4 and 5 in][]{Westmeier2021}. In particular, the noise term is obtained by scaling the noise of the cube by the number of independent voxels included in the 3D detection mask. While the \HI calibration uncertainty is factored into the total \MHI error, the detectability of an \HI source only depends on the statistical error. The number of all \HI-sources (black outline histogram) in each bin increases until log(\MHI/\Msol) = 6, where the statistical SNR is $\sim$ 5. We are therefore able to confidently probe the shape of the HIMF low-mass slope down to  log(\MHI/\Msol) = 6. The number of detections increases as expected, with a total of 49 below log(\MHI/\Msol) = 7. However, only five of those are galaxies, demonstrating a clear break in the distribution of galaxies at log(\MHI/\Msol) = 7, where the typical statistical SNR of the \HI-sources is $\sim$ 20 (Fig.\ref{fig:HI_SNR}). This abrupt truncation occurs at such a high SNR that it cannot be attributed to \HI-bearing galaxies being lost in the noise. 

To assess the completeness, we implement the modified Rauzy test \citep{Rauzy2001}. Unlike the original Rauzy test designed for flux-limited optical galaxy catalogues \citep{Rauzy2001}, the modified version accounts for non-uniform flux sensitivity and the variation of velocity widths in \HI sources. It provides a statistical estimate of the flux down to which the \HI catalogue is complete, without requiring assumptions about the shape of the \HIMF. The modified Rauzy test has been employed in previous \HIMF catalogues \citep[e.g.][]{Zwaan2004, Toribio2011, Staveley-Smith2016, Said2019, Xi2021}, and relies on the completeness variable \Tc, a test statistic derived from the cumulative, volume-weighted distribution of sources. In a statistically complete, flux-limited sample, the expected distribution of ranked sources follows a uniform trend; deviations from this trend cause \Tc to fall. When \Tc steadily decreases towards lower flux values, the sample becomes incomplete at the intersection where \Tc = $-2$ or $-3$, corresponding to 97.7\% or 99.4\% confidence levels, respectively. \Tc, which quantifies deviations in the cumulative, volume-weighted distribution of sources from the uniform distribution expected in a statistically complete, flux-limited survey. For each galaxy, its maximum detectable volume (V$_{\mathrm{max}}$) is estimated using its integrated flux and velocity width, and \Tc is calculated from the cumulative sum of 1/V$_{\mathrm{max}}$-weighted ranks. Under completeness, the ranked distribution follows a uniform trend and \Tc follows a normal distribution with mean zero. A systematic deficit of low-flux sources causes \Tc to drift negative values.

For the computation of \MHI in Table\,\ref{tab:gal_detections}, we adopt the nominal distance of the cluster at 20\,Mpc \citep{Blakeslee2009}. However, the decrease in \Tc relies on the \HI-sources being in a volume i.e. located at different distances, and using $D$ = 20\,Mpc for all sources would not produce a meaningful \Tc. As the true distance of the \HI sources is not known, we run a Monte Carlo-style simulation where we assign a random distance within $D$ = 20 $\pm$ 2.1\,Mpc (3\Rvir) to each \HI source and compute 1000 realisations of \Tc. The variation in distances alters the observed number of \HI sources with that flux at that distance compared to the expected distribution, and therefore provides a meaningful \Tc, along with a more realistic \HI source distribution in the Fornax cluster. 

\begin{figure}
    \centering
    \includegraphics[width=1\linewidth]{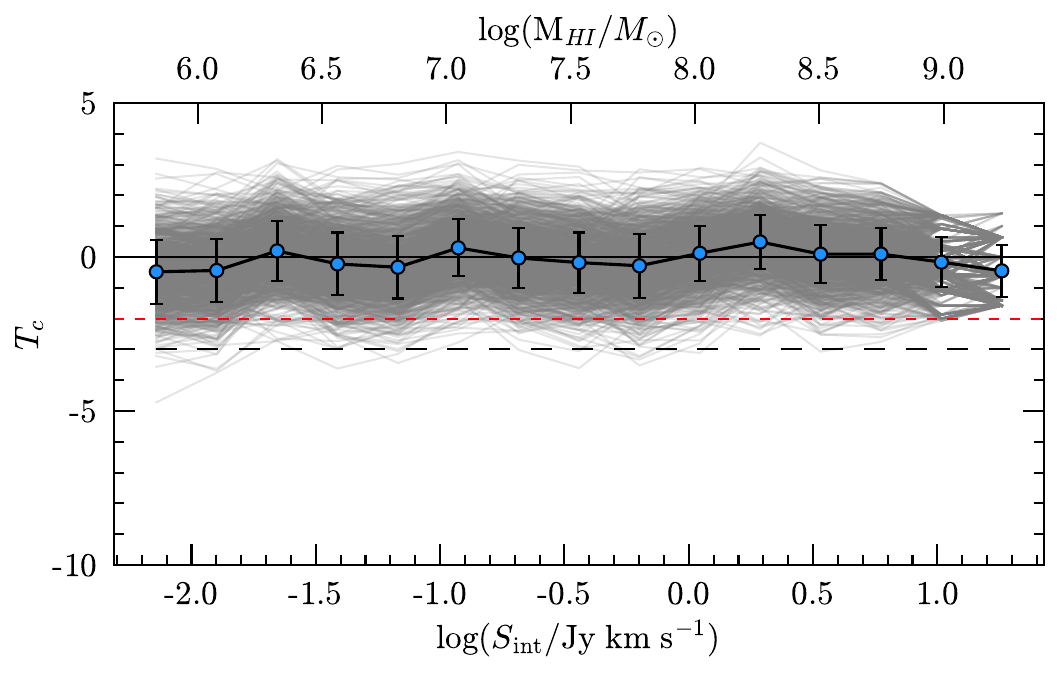}
    \caption{The modified Rauzy test. The Rauzy completeness statistic (\Tc) as a function of integrated flux. The blue points represent $\langle T_{c} \rangle$ from 1000 runs varying the line-of-sight distance of each detection within 3\Rvir of the Fornax cluster, the uncertainties are computed as the variance and the grey lines show each individual test. The red, short-dashed and black, long-dashed lines represent the 97.7\% (\Tc = –2) and 99.4\% (\Tc = –3) confidence intervals.}
    \label{fig:Rauzy}
\end{figure}

In Fig.\,\ref{fig:Rauzy}, we show $\langle T_{c} \rangle$ as a function of integrated flux, computed from the 1000 realisations. The individual \Tc values for each realisation are also shown in order to highlight the scatter. $\langle T_{c} \rangle$ is correlated with the number of all \HI sources (black histogram in Fig.\,\ref{fig:HI_SNR}), although it is worth noting there is no \Tc value for NGC\,1365 as it is the only source in the brightest bin. Although within the uncertainty, $\langle T_{c} \rangle$ starts to decrease below $S_\mathrm{int}$ $\sim$ 1.8 $\times$ 10$^{-2}$\,Jy\,\kms (equivalent to log(\MHI/\Msol) $\sim$ 6.2), hinting at the onset of incompleteness; this is consistent with the distribution of \HI sources in Fig.\,\ref{fig:HI_SNR}). As $\langle T_{c} \rangle$ never crosses $-2$ or $-3$, we are not missing a significant number of \HI sources, and we are able to robustly probe the shape of the \HIMF of galaxies down to log(\MHI/\Msol) = 6.0. 

The main caveats to take into consideration, is that the Rauzy statistic is designed to assess completeness on a flux-limited sample in larger volumes than a single cluster. Despite our sample not being purely flux-limited, the test is still valid by using the line-widths to compute the maximum detectable volume for each source in every distance realisation. This assesses whether there is any flux-based incompleteness relative to the survey effective selection function. We further test the behaviour of \Tc by injecting fake sources into our \HI catalogue in a variety of volumes, and were able to determine that an additional low-mass (e.g. log(\MHI/\Msol) $<$ 5.7) source along with a larger (by a factor $\sim$ 5) volume caused $\langle T_{c} \rangle$ to drop below $-3$ at $S_\mathrm{int}$ $\sim$ 3.7 $\times$ 10$^{-2}$\,Jy\,\kms -- equivalent to log(\MHI/\Msol) $\sim$ 5.5, indicating statistical incompleteness. This is the expected behaviour of \Tc and happens because the larger volume increases the maximum detectable volume for many sources, reducing their contribution to the cumulative, volume-weighted rank distribution. If no additional detections are present in the increased volume, the ranked distribution of sources deviates significantly from the uniform expectation, driving $\langle T_{c} \rangle$ to lower values.

There are 54 cluster galaxies brighter than $m_{r} = 16$ with optical redshifts and no \HI detection, and as a final test, we make use of spectral stacking to search for missed \HI in cluster galaxies below the noise. We extract 800\,\kms-wide \HI spectra centred on the position and optical redshift of each galaxy. Then, we stack the galaxies in six sub-samples defined by colour cuts starting at $g - r$ = 0.5 and increasing by 0.1 increments. There is no \HI signal in any of the stacked spectra, which lends further support to the idea that we are not missing a large number of galaxies with \HI, and on average, there is no low level detectable \HI in galaxies with known redshifts, irrespective of their colour.

\section{The Fornax cluster HIMF}
\label{sec:HIMF}
It is immediately obvious from Fig.\,\ref{fig:HI_SNR} that there is a truncation in the \HI-detected galaxy number density below log(\MHI/\Msol) = 7. A hint of this was first noted in a Bayesian analysis on the Fornax \HIMF using Australia Telescope Compact Array (ATCA) data \citep{Loni2022}. As the MFS is much more sensitive than the ATCA observations, we are now able to definitively quantify the shape of the \HIMF in the low-mass regime. 

By definition, a Schechter function is not going to fit the full range of \MHI values, particularly below log(\MHI/\Msol) = 7. Therefore, we make use of the modified maximum likelihood (MML) method implemented in \texttt{dftools} \citep{Obreschkow2018} to fit a Schechter function over the range of \MHI values that follow a power-law i.e. above log(\MHI/\Msol) = 7\footnote{When fitting the full \MHI range, the MML diverged and was unable to fit a Schechter function, confirming that the \MHI distribution does not follow the same power-law above log(\MHI/\Msol) = 7.}. When a unique solution exists for the maximum likelihood estimator, the MML method has demonstrated to be a analytical approximation of Bayesian hierarchical modelling, and has shown good agreement with the 1/V$_{max}$ method \citep[e.g.][]{Westmeier2017, Obreschkow2018, Ponomareva2023, Yu2025}. In comparison, the MML method has the distinct advantage of not binning the data and recovering the mass function, even with small number statistics. This is important for the Fornax \HIMF as we are limited to a small volume with small number statistics. 

The MML estimator can be biased (i.e. the expectation of the sample differs from the true population) and is particularly problematic when the observed mass function only has a few points. \citet{Obreschkow2018} showed that with N = 5, there are significant errors in the parameter estimation. When N $\leq$ 10$^{3}$ there is large scatter during the fitting, however, the recovery of the true mass function works remarkably well for N $\geq$ 10 when accounting for the MML estimator bias through jackknifing -- where the fit is repeated N times with each data point omitted once, and the resulting variation is used to estimate the bias and uncertainty of the fit parameters \citep{Obreschkow2018}. We fit our \HI catalogue with N = 30 (for log(\MHI/\Msol) $\geq$ 7) using the MML method, and jackknife the sample to ensure good recovery of the true mass function. Our catalogue spans a large \MHI range, which is advantageous for constraining the shape of the low-mass end, the main goal of this work.  

\begin{figure*}
    \sidecaption
    \includegraphics[width=12cm]{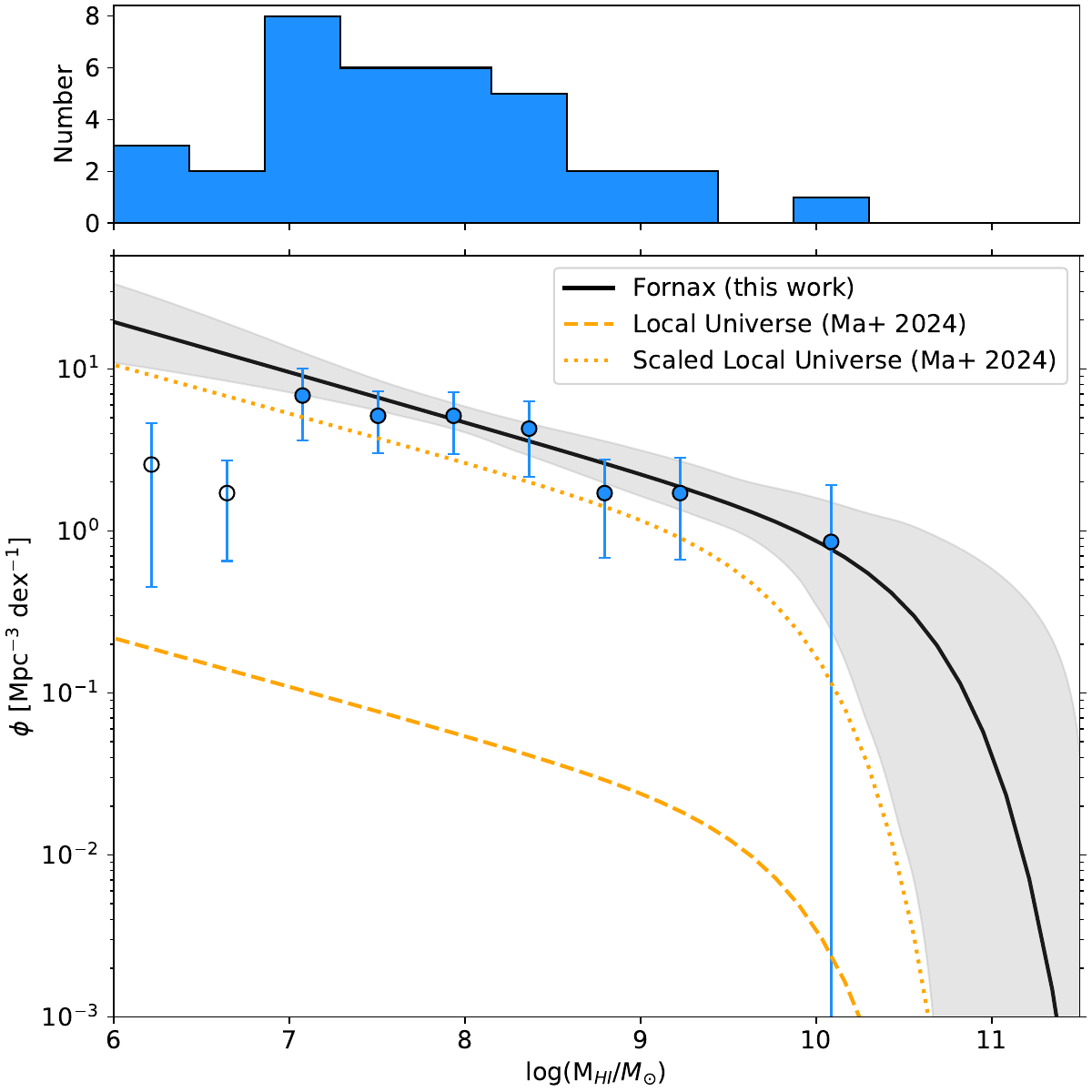}
    \caption{Fornax Cluster \HIMF. In the main figure, the circles represent the observed number density ($\phi$) from the binned data (for display purposes) with their respective Poisson uncertainties. The filled circles represent galaxies in the \MHI range used to fit the \HIMF, log(\MHI/\Msol) $\geq$ 7, while the open circles represent galaxies in the \MHI range excluded from the fit because of the obvious departure from a power-law. The black line is the MML fit and the grey shaded region denotes the 1$\sigma$ uncertainty of the fit. A volume of 2.7\,Mpc$^{3}$ was assumed for the fitting procedure. The dashed orange line shows the \HIMF of the local Universe \citep{Ma2025}, and the dotted orange line is the local Universe \HIMF scaled by using the Fornax $\phi_{*}$. The top panel contains the histogram of the \HI-detected galaxies.}
    \label{fig:HIMF}
\end{figure*}

We present the Fornax cluster \HIMF in Fig.\,\ref{fig:HIMF}, the deepest \HIMF to date, outside the Local Group. The measured low-mass slope and characteristic knee mass are $\alpha$ = $-1.31$ $\pm$ 0.13, log(M$_{*}$/\Msol) = 10.52 $\pm$ 1.89, in an assumed volume\footnote{Calculated by converting the mosaic’s angular area to a physical area, assuming a depth of 2\Rvir and using a scale of 1\arcmin = 5.79\,kpc.}  of 2.7\,Mpc$^{3}$. Mass bins with low counts (e.g. a single entry) are statistically dominated by uncertainties, rendering them unconstrained. This is true for the knee at the high-mass end, as there is just a single galaxy (NGC\,1365) that has a log(\MHI/\Msol) $>$ 9.5. This is not true for the low-mass end, despite the bin 6.43 $<$ log(\MHI/\Msol) $\leq$ 6.86 only containing a single galaxy. The difference can be seen in Fig.\,\ref{fig:Mr_MHI}, demonstrating that in the Fornax cluster, there are very few bright (or massive) galaxies with \HI, while on the other hand, there are hundreds of low-mass galaxies that could host \HI in the 6.43 $<$ log(\MHI/\Msol) $\leq$ 6.86 range. The normalisation constant $\phi_{*}$ is a volume scaling factor for the \HIMF, has no impact on the shape of the low-mass end and should not be directly compared to other H\textsc{i}MFs. For these reasons, we only discuss the shape of the low-mass end and its implications in the rest of the paper.

The abrupt truncation previously noted in the distribution of \HI-detected galaxies at log(\MHI/\Msol) = 7 (Fig.\,\ref{fig:HI_SNR}) directly results in the shape of the observed \HIMF to depart from a Schechter function (Fig.\,\ref{fig:HIMF}). As discussed in detail in the previous section, there is strong evidence that the departure is physical and not due to catalogue incompleteness. Tens of additional galaxies with \HI would be required in the 6 $<$ log(\MHI/\Msol) $\leq$ 7 range for the observed number density to follow a power-law. The typical SNR is $\sim$ 5 and 20 for \HI-sources with log(\MHI/\Msol) = 6 and 7, respectively (Fig.\,\ref{fig:HI_SNR}), making it extremely unlikely that we have missed such a large number of galaxies at this sensitivity, and this \MHI range. Furthermore, we detect the majority of \HI clouds in this range, which demonstrates that we would have detected the galaxies if they were there. The Rauzy test additionally supports that there is a physical absence of \HI-detected galaxies in this mass range and the departure from a Schechter function is not caused by catalogue incompleteness. 

\subsection{The make up and evolution of the \HIMF}
\label{sec:evolution}
The \HIMF is an important tool to study galaxies because, due to the large scatter in \MHI/\Mstar at fixed \Mstar \citep[][]{Saintonge2022}, galaxies contributing to similar parts of the luminosity or stellar mass function can populate very different parts of the \HIMF. This distinction is evident when comparing the two optically brightest \HI detections in our catalogue (Table\,\ref{tab:gal_detections}). NGC\,1365, the most luminous and massive \HI-detected galaxy, has $M_{r}$ = $-22.46$ and log(\MHI/\Msol) = 10.2. It has a typical \MHI for a star-forming galaxy and defines the knee of the Fornax \HIMF. In contrast, NGC\,1387, the second brightest optical galaxy in the \HI catalogue, has $M_{r}$ = $-21.32$ but a much lower log(\MHI/\Msol) = 6.3, placing it in the lowest mass bin of the \HIMF.

In Fig.\,\ref{fig:Mr_MHI}, the majority (63\%) of the galaxies that make up the \HIMF lie within the scatter (grey-shaded region) of the $M_{r}$-\MHI scaling relation. There are five galaxies just below that scatter, and six galaxies (17\%) substantially below the relation. Therefore, 83\% of the galaxies contain the expected (or close to) amount of \HI given its luminosity. In Fornax the faint galaxies are contributing to the low-mass regime of the \HIMF, and it is a rare occurrence for a bright galaxy to contribute to the same part of the \HIMF. This is unsurprising given the galaxy population in the Fornax cluster is dominated by low-mass galaxies (histogram in Fig.\,\ref{fig:Mr_MHI}). 

Another important consideration is how the \HIMF evolves over time. As galaxies lose \HI, they move to lower \HI masses along the \HIMF. The degree to which the shape of the low-mass slope can change depends on the galaxy mass distribution and efficiency of \HI stripping. \citet{Kleiner2023} showed that low-mass galaxies experience rapid \HI stripping in the Fornax cluster, which is the likely to be the main cause of the collapse of the \HIMF. Since there are so few galaxies at the high-\MHI end of the \HIMF (Figs.\,\ref{fig:HI_SNR} and \ref{fig:HIMF}), even if they lose a substantial amount of \HI, they are not enough to make up for the loss of faint galaxies from the low-\MHI end of the \HIMF.

\section{Comparison to other H\textsc{i}MFs}
\label{sec:comparison}
Recently, \citet{Ma2025} measured the most complete \HIMF at $z$ $\sim$ 0 using FASHI \citep{Zhang2024} in combination with the HIPASS \citep{Zwaan2005} and ALFALFA catalogues \citep{Jones2018}. This \HIMF covered 76\% of the entire sky, which minimises cosmic variance and is a typical representation of the field. The low-mass slope was measured to be $\alpha$ = $-1.3$ and agrees with previous measurements of field H\textsc{i}MFs. The \citet{Ma2025} \HIMF measured the low-mass slope down to log(\MHI/\Msol) = 7, making it the deepest field HIMF to date, and we compare their log(M$_{*}$/\Msol) = 9.86, $\alpha$ =  $-1.3$, and $\phi_{*}$ = 6.58 $\times$ 10$^{-3}$\,Mpc$^{-3}$\,dex$^{-1}$, to the Fornax \HIMF in Fig.\,\ref{fig:HIMF}. The Fornax observed $\phi$ clearly represents an over-density of \HI-detected galaxies compared to the field. Remarkably, the Schechter function low-mass slope of $\alpha$ = $-1.31$ matches the field \HIMF \citep{Ma2025}, for log(\MHI/\Msol) $\geq$ 7. 

The fact that the Fornax \HIMF has the same slope as the field can be attributed to the similarities in the galaxy population down to log(\MHI/\Msol) = 7. Even though the environments are different, and we observe the effects of active \HI removal (Fig.\,\ref{fig:MFS_overlay}) in the Fornax cluster \citep[e.g.]{Kleiner2023, Serra2023, Serra2024}, as discussed in Section\,\ref{sec:evolution} and shown in Fig.\,\ref{fig:Mr_MHI} the majority of the \HI-detected Fornax galaxies have the expected (or close to) amount of \HI as field galaxies of the same luminosity \citep{Kovac2009}. While there is ubiquitous evidence that the \HI content of Fornax galaxies is being affected by tidal fields and ram pressure \citep[e.g.][]{Serra2023}, most of the \HI-bearing galaxies in Fornax must have only recently joined the cluster, otherwise cluster processes would have removed a more substantial amount of \HI \citep{Loni2021}. 

The low-mass end of the \HIMF is the most sensitive to environmental effects and there has been contentious debate about the shape of it, particularly below log(\MHI/\Msol) = 8. While some H\textsc{i}MFs \citep[e.g.][]{Verheijen2001, Zwaan2005, Kovac2007, Freeland2009, Stierwalt2009, Martin2010, Pisano2011, Westmeier2017, Jones2018, Busekool2021} include direct detections of galaxies with \HI below log(\MHI/\Msol) = 8, all of them apply significant completeness corrections, that make the shape of the low-mass end the most uncertain. Many of these studies are dominated by the uncertainties from low-number statistics coupled with completeness correction in this mass range, which has made accurately constraining the shape of the low-mass end an elusive task. Out of these studies, only a few \citep[e.g][]{Kovac2007, Stierwalt2009, Martin2010, Westmeier2017, Jones2018} probe just below log(\MHI/\Msol) = 7, where the shape of the \HIMF is the least constrained. 

In the higher density environments, a variety of low-mass slopes have been measured. Multiple studies have reported a flat ($-1.3$ $\leq$ $\alpha$ $<$ $-1$) slope in the group environment \citep[e.g.][]{Verheijen2001, Freeland2009, Kilborn2009, Pisano2011, Westmeier2017, Said2019, Jones2020, Yu2025} where they attribute the flattening to \HI removal from low mass galaxies. Other studies do not measure any noticeable flattening with a slope the same as the field ($\alpha$ $\sim$ $-1.3$), and report no environmental dependence on the low-mass slope \citep[e.g][]{Martin2010, Jones2016}. There are also a small number of galaxy group studies that measure a slope steeper ($\alpha$ $<$ $-1.4$) than the field \citep{Stierwalt2009, Davies2011}, and claim the difference is a result of cosmic variance or a population of gas-rich dwarfs not present or detectable in other H\textsc{i}MFs. 

Our Fornax result is different from those reported above for galaxy groups in that we do not report a decrease of the \HIMF slope, but rather a normal slope followed by the collapse of the \HIMF below log(\MHI/\Msol) = 7. This sudden change in $\phi$ is abrupt and fitting a Schechter function below log(\MHI/\Msol) = 7 neither describes nor fits the observed number density. The flat H\textsc{i}MFs in the literature may have tentatively observed a similar effect, albeit restricted by sensitivity limitations and large completeness corrections. The Fornax \HIMF is the first robust measurement of a departure from a Schechter function\footnote{Incidentally, including the \HI clouds (Table\,\ref{tab:cloud_detections}) in our analysis would make the \HIMF consistent with that of the field down to log(\MHI/\Msol) = 6 . However, since we have established that the clouds are most definitely not galaxies (Fig\,\ref{fig:Mr_MHI}) and most are associated with gas removal from the two \HI-richest galaxies in the cluster (Fig\,\ref{fig:MFS_overlay}), we do not think that this result is relevant for our conclusion on the collapse of the \HIMF of galaxies in Fornax.} not hindered by incompleteness, and is directly related to the efficient removal of \HI from low-mass galaxies \citep[described in][]{Kleiner2023}. 

We expect that this phenomenon is not unique to Fornax, and that it is a common occurrence in clusters. However, whether the \HIMF deviates from a Schechter function, and at what mass this occurs, depends on the \HI-galaxy population and the efficiency of environmental \HI-stripping from galaxies. We therefore hypothesize that the collapse of the \HIMF is correlated with environmental density i.e. the \HIMF collapses at a higher value of \MHI in higher-density environments. Indeed, the Coma cluster is more massive than Fornax, and by using spectral stacking, \citet{Healy2021} showed that Coma galaxies are either directly detected in \HI at log(\MHI/\Msol) $>$ 8 or contain orders of magnitude less \HI than expected. This bimodal \MHI distribution implies rapid gas removal and quenching, where the drop happens at log(\MHI/\Msol) = 8, one dex higher than in Fornax. Based on our findings, we would therefore expect the low-mass end of the Coma \HIMF to also significantly deviate from a Schechter function, but at a higher \MHI compared to Fornax. The fact that, while collapsing below log(\MHI/\Msol) = 7, the Fornax galaxy \HIMF has the same slope as the field \HIMF above log(\MHI/\Msol) = 7 may tell us something about the timescale of gas removal. We hypothesize that \HI-rich galaxies have only recently arrived in Fornax, and that there has only been enough time to strip the smallest galaxies of their \HI, causing the observed \HIMF collapse. More massive galaxies may be losing gas at a sufficiently slow rate that the \HIMF slope has not changed significantly above log(\MHI/\Msol) = 7 yet. Given more time and no substantial accretion of new \HI-rich members, it is possible that the \HIMF of Fornax will also start to flatten, as observed in other over-densities. In summary, our hypothesis is that in clusters: a sharp \HIMF cut-off develops fairly quickly; more time is needed to change the slope of the \HIMF above the cut-off; and both the cut-off mass and the speed at which the \HIMF slope changes above the cut-off depend on cluster mass.

\section{Summary}
\label{sec:conclusion}
We present the deepest \HI mass function of galaxies ever measured, outside the Local Group. Using observations from the MeerKAT Fornax Survey, we detect \HI in 35 galaxies and 44 clouds with no optical counterparts within the Fornax cluster $\sim$ \Rvir. Using the deep optical images from the Fornax Deep Survey, we show a $\sim$ four magnitude gap between the 5$\sigma$ optical flux limit of the \HI clouds and \HI-detected galaxies in the $M_{r}$-\MHI plane. We conclude that the \HI clouds are a distinct population of \HI-rich, optically ultra-faint objects (undetected in the FDS data) rather than a smooth continuation of the population of \HI-detected galaxies towards progressively lower luminosities. We show that 33 out of the 44 \HI clouds are associated with the two galaxies with the highest \MHI in the cluster -- NGC\,1427A and NGC\,1365. However, the clouds themselves contribute a negligible amount of mass to the total \MHI budget of either galaxy. 

We show that the number count of \HI-sources increases as expected down to log(\MHI/\Msol) = 6. The typical SNR of \HI sources with log(\MHI/\Msol) = 6 and 7 is $\sim$ 5 and 20, respectively, demonstrating that we can robustly measure the shape of the \HIMF down to log(\MHI/\Msol) = 6. There is an abrupt truncation in the distribution of \HI-detected galaxies at log(\MHI/\Msol) = 7, which is physical and not a result of missed detections. We estimate the Rauzy completeness statistic by running a Monte Carlo simulation to account for the uncertainty in distance to each \HI source. We note a tentative hint of the start of catalogue incompleteness below log(\MHI/\Msol) = 6.2, although not enough to be statistically significant, which further supports we are detecting most, if not all \HI sources above log(\MHI/\Msol) = 6. 

The sudden drop in \HI-detected galaxies below log(\MHI/\Msol) = 7 prevents a Schechter function from adequately fitting the full \MHI range. We therefore only fit a Schechter function, using the modified maximum likelihood method and jackknife the sample to accurately recover the underlying mass function of galaxies in the \MHI range that follows a power-law -- down to log(\MHI/\Msol) = 7. The measured low-mass slope is $\alpha$ = $-1.31$ $\pm$ 0.13, matching the low-mass slope of field H\textsc{i}MFs. Below log(\MHI/\Msol) = 7, the galaxy \HIMF collapses, due to a lack of detections in the 6 $<$ log(\MHI/\Msol) $\leq$ 7 range. The abrupt departure from the Schechter function at the low-mass end of the galaxy \HIMF is not due to a lack of sensitivity, but rather a real absence of low \MHI galaxies. There would need to be tens of an additional \HI-detected galaxies in this \MHI range to follow the same power-law above log(\MHI/\Msol) = 7.

An observed collapse of the \HIMF gives direct insight into the efficiency of \HI removal and the population of \HI-detected galaxies. The more efficient the \HI removal is, the higher the \MHI where the collapse will occur. The shape of the Fornax \HIMF at the low-mass end is directly related to the efficient removal of \HI from low-mass (e.g. $M_{r}$ $>$ $-14$) cluster galaxies. Even though the low-mass end of the \HI-detected galaxy number density can be re-populated by massive galaxies containing small amounts of \HI, there are too few massive galaxies in the cluster to alter the departure from a Schechter function, observed at log(\MHI/\Msol) = 7. Previous studies have noted flat slopes of $\alpha$ $\sim$ $-1$ in groups, which may be tentative evidence of what we observe in the Fornax cluster. However, this is the first robust measurement an abrupt departure from a Schechter function in the \HIMF low-mass regime.

With deep, resolved \HI studies from MeerKAT and the upcoming Square Kilometre Array (SKA), we expect the collapse of the \HIMF to be observed in future studies.

\begin{acknowledgements}
The MeerKAT telescope is operated by the South African Radio Astronomy Observatory, which is a facility of the National Research Foundation, an agency of the Department of Science and Innovation. We are grateful to the full MeerKAT team at SARAO for their work on building and commissioning MeerKAT. This work made use of the Inter-University Institute for Data Intensive Astronomy (IDIA) visualisation lab (https://vislab.idia.ac.za). IDIA is a partnership of the University of Cape Town, the University of Pretoria and the University of Western Cape.

This work has received funding from the European Research Council (ERC) under the European Union’s Horizon 2020 research and innovation programme (grant agreement No. 882793 "MeerGas"). This project has received funding from the European Research Council (ERC) under the European Union’s Horizon 2020 research and innovation programme (grant agreement no. 679627; project name FORNAX).

The data of the MeerKAT Fornax Survey are reduced using the CARACal pipeline, partially supported by ERC Starting grant number 679627, MAECI (Italian Ministry of Foreign Affairs and International Cooperation) Grant Numbers PGR ZA23GR03 (RADIOMAP) and PGR ZA18GR02 (RADIOSKY2020), DSTNRF Grant Number 113121 as part of the ISARP Joint Research Scheme, and BMBF project 05A17PC2 for D-MeerKAT.

FMM carried out part of the research activities described in this paper with contribution of the Next Generation EU funds within the National Recovery and Resilience Plan (PNRR), Mission 4 -- Education and Research, Component 2 -- From Research to Business (M4C2), Investment Line 3.1 -- Strengthening and creation of Research Infrastructures, Project IR0000034 -- ``STILES - Strengthening the Italian Leadership in ELT and SKA''. SHAR and RCKK acknowledge the support by the South African Research Chair initiative by the DST and NRF. PK is partially supported by the BMBF project 05A23PC1 for D-MeerKAT. 

\end{acknowledgements}

\bibliographystyle{aa} 
\bibliography{References} 

\begin{appendix} 
\begin{table*}
    \centering
    \caption{The \HI detected galaxies.}
    \begin{tabular}{c c c c c c }
        \hline
        \hline
        Name & FCC Number & log(\MHI/\Msol) & $v_{\rm sys}$ & $w_{20}$ & M$_{r}$ \\
             &            &                 & (\kms)        & (\kms)   & (mag)   \\
        (1)  & (2)        & (3)             & (4)           & (5)      & (6)     \\    
        \hline
        ESO\,358-015      & 113   & 8.2$\substack{+0.04 \\ -0.05}$  & 1388 & 98 & $-17.03$  \\
        ESO\,358-016      & 115   & 8.1$\substack{+0.04 \\ -0.05}$  & 1700 & 95 & $-15.43$  \\
        ESO\,358-051      & 263   & 8.4$\substack{+0.04 \\ -0.05}$  & 1724 & 142 & $-18.33$  \\
        ESO\,358-063      & 312   & 9.3$\substack{+0.04 \\ -0.05}$  & 1929 & 311 & $-20.02$  \\
        ESO\,358-064      & B1899 & 7.7$\substack{+0.04 \\ -0.05}$  & 1465 & 127 & $-16.20$  \\
        ESO\,359-002      & 335   & 7.3$\substack{+0.05 \\ -0.05}$  & 1447 & 95 & $-18.18$  \\ 
        FCC\,057          & 57    & 7.2$\substack{+0.05 \\ -0.05}$  & 1442 & 59 & $-14.37$  \\ 
        FCC\,090          & 90    & 7.7$\substack{+0.04 \\ -0.05}$  & 1825 & 79 & $-17.60$  \\ 
        FCC\,102          & 102   & 8.0$\substack{+0.04 \\ -0.05}$  & 1640 & 59 & $-15.80$  \\
        FCC\,120          & 120   & 8.0$\substack{+0.04 \\ -0.05}$  & 846 & 45 & $-15.72$  \\
        FCC\,128          & 128   & 7.1$\substack{+0.04 \\ -0.05}$  & 1543 & 50 & $-15.50$  \\
        FCC\,134          & 134   & 6.4$\substack{+0.06 \\ -0.07}$  & 1395 & 35 & $-14.65$  \\ 
        FCC\,207          & 207   & 6.1$\substack{+0.07 \\ -0.08}$  & 1420 & 43 & $-16.55$  \\
        FCC\,248          & 248   & 6.1$\substack{+0.07 \\ -0.09}$  & 1859 & 15 & $-13.44$  \\
        FCC\,261          & 261   & 7.2$\substack{+0.05 \\ -0.05}$  & 1665 & 62 & $-16.44$  \\ 
        FCC\,267          & 267   & 7.2$\substack{+0.05 \\ -0.05}$  & 833 & 50 & $-16.16$  \\ 
        FCC\,282          & 282   & 7.1$\substack{+0.05 \\ -0.05}$  & 1260 & 69 & $-17.79$  \\ 
        FCC\,299          & 299   & 7.5$\substack{+0.04 \\ -0.05}$  & 2100 & 104 & $-15.18$  \\
        FCC\,306          & 306   & 8.3$\substack{+0.04\\ -0.05}$   & 884 & 130 & $-16.00$  \\ 
        FCC\,323          & 323   & 7.6$\substack{+0.04 \\ -0.05}$  & 1499 & 47 & $-14.41$  \\
        FCC\,332          & 332   & 7.2$\substack{+0.05 \\ -0.05}$  & 1325 & 118 & $-14.41$  \\ 
        FCCB\,0507        & B507  & 7.8$\substack{+0.04 \\ -0.05}$  & 1149 & 74 & $-13.87$  \\ 
        FCCB\,0905        & B905  & 7.7$\substack{+0.04 \\ -0.05}$  & 1246 & 70 & $-15.12$  \\ 
        FCCB\,1461        & B1461 & 7.1$\substack{+0.05 \\ -0.05}$  & 799 & 34 & $-13.58$  \\ 
        FDS6\,DWARF\,100b & --    & 7.1$\substack{+0.04 \\ -0.05}$  & 1983 & 36 & $-13.06$  \\ 
        FDS14\,DWARF\,011 & --    & 6.5$\substack{+0.10 \\ -0.12}$  & 1430 & 48 & $-12.73$  \\ 
        NGC\,1351A        & 67    & 8.8$\substack{+0.04 \\ -0.05}$  & 1347 & 221 & $-18.48$  \\ 
        NGC\,1365         & 121   & 10.2$\substack{+0.04 \\ -0.05}$ & 1636 & 400 & $-22.46$  \\ 
        NGC\,1386         & 179   & 7.7$\substack{+0.04 \\ -0.05}$  & 868 & 48 & $-20.67$  \\ 
        NGC\,1387         & 184   & 6.6$\substack{+0.07 \\ -0.08}$  & 1302 & 155 & $-21.32$  \\  
        NGC\,1427         & 276   & 7.6$\substack{+0.04 \\ -0.05}$  & 1390 & 110 & $-21.05$  \\ 
        NGC\,1427A        & 235   & 9.4$\substack{+0.04 \\ -0.05}$  & 2036 & 120 & $-18.75$  \\
        NGC\,1436         & 290   & 8.3$\substack{+0.04 \\ -0.05}$  & 1395 & 225 & $-20.23$  \\
        NGC\,1437A        & 285   & 8.8$\substack{+0.04 \\ -0.05}$  & 886 & 93 & $-18.02$  \\
        NGC\,1437B        & 308   & 8.5$\substack{+0.04 \\ -0.05}$  & 1510 & 142 & $-18.67$  \\        
        \hline 
        \hline
    \end{tabular}
    \\
    \textbf{Notes.} (1) Galaxy name. (2) FCC number. (3) \MHI and its uncertainty as measured from the 41\arcsec-resolution image. (4) The systemic velocity measured in the optical definition. (5) Width of the \HI profile at 20\% of the peak flux, from the \SoFiA parametrisation. (6) The absolute $r$-band magnitude derived from aperture photometry from \citet{Su2021}, assuming a distance of 20\,Mpc. 

    \label{tab:gal_detections}
\end{table*}

\begin{table*}
    \centering
    \caption{Detected \HI clouds, with their position and nearest cluster galaxy.}
    \begin{tabular}{c c c c c c c}
        \hline
        \hline
        Nearest galaxy & RA (J2000) & Dec (J2000) & log(\MHI/\Msol) & $v_{\rm sys}$ & $w_{20}$ & $\lvert\Delta v_{\rm sys}\rvert$ \\
                       & (hms)      & (dms)       &                 & (\kms)        & (\kms)   & (\kms)     \\
        (1)            & (2)        & (3)         & (4)             &  (5)          & (6)      & (7)        \\
        \hline
        ESO\,358-015   & 03:32:32.2	& -34:42:16 & 6.6$\substack{+0.06 \\ -0.07}$ & 1176 & 64 & 212 $^{(a)}$ \\
        ESO\,358-016   & 03:33:54.9 & -35:39:01 & 6.1$\substack{+0.09 \\ -0.11}$ & 1245 & 38 & 455 $^{(a)}$ \\
        ESO\,358-063   & 03:45:10.6 & -34:40:09 & 6.0$\substack{+0.08 \\ -0.10}$ & 1251 & 27 & 678 $^{(a)}$ \\
        FCC\,306       & 03:45:17.2	& -36:32:45 & 6.4$\substack{+0.07 \\ -0.08}$ & 925  & 49 & 39 $^{(a)}$  \\  
        NGC\,1351A     & 03:27:51.8	& -35:30:01 & 6.3$\substack{+0.08 \\ -0.09}$ & 1549 & 40 & 202 $^{(a)}$ \\ 
        NGC\,1399      & 03:37:52.5	& -35:14:15 & 5.9$\substack{+0.1  \\ -0.13}$ & 1701 & 26 & 276 $^{(b)}$ \\
        NGC\,1399      & 03:38:10.3	& -35:16:33 & 6.2$\substack{+0.09 \\ -0.11}$ & 1210 & 42 & 215 $^{(b)}$ \\
        NGC\,1399      & 03:38:25.3 & -35:19:32 & 6.6$\substack{+0.07 \\ -0.08}$ & 1548 & 50 & 123 $^{(b)}$ \\
        NGC\,1427      & 03:43:48.5	& -35:20:38 & 6.1$\substack{+0.08 \\ -0.11}$ & 1575 & 25 & 187 $^{(a)}$ \\ 
        NGC\,1427A     & 03:39:42.6 & -35:47:45 & 6.4$\substack{+0.07 \\ -0.09}$ & 1769 & 52 & 267 $^{(a)}$ \\ 
        NGC\,1427A     & 03:40:04.0 & -36:05:52 & 6.2$\substack{+0.07 \\ -0.09}$ & 1943 & 38 & 93 $^{(a)}$  \\ 
        NGC\,1427A     & 03:40:56.8 & -36:27:10 & 6.1$\substack{+0.08  \\ -0.1}$ & 1811 & 17 & 225 $^{(a)}$ \\
        NGC\,1427A     & 03:41:02.6	& -36:01:50 & 6.3$\substack{+0.07 \\ -0.09}$ & 1776 & 27 & 260 $^{(a)}$ \\
        NGC\,1427A     & 03:41:35.9 & -36:22:59 & 6.3$\substack{+0.07 \\ -0.08}$ & 1791 & 30 & 245 $^{(a)}$ \\
        NGC\,1427A     & 03:41:37.2 & -36:27:01 & 6.3$\substack{+0.07 \\ -0.09}$ & 1773 & 44 & 263 $^{(a)}$ \\
        NGC\,1427A     & 03:41:52.2 & -36:13:56 & 6.3$\substack{+0.07 \\ -0.09}$ & 1801 & 29 & 235 $^{(a)}$ \\
        NGC\,1427A     & 03:42:05.5 & -36:25:32 & 6.1$\substack{+0.09 \\ -0.11}$ & 1772 & 33 & 264 $^{(a)}$ \\
        NGC\,1427A     & 03:42:13.4 & -36:10:56 & 6.5$\substack{+0.07 \\ -0.08}$ & 1772 & -- & 264 $^{(a)}$ \\
        NGC\,1427A     & 03:42:15.7 & -36:09:39 & 6.2$\substack{+0.08  \\ -0.1}$ & 1772 & -- & 264 $^{(a)}$ \\
        NGC\,1427A     & 03:42:19.1 & -36:15:35 & 6.3$\substack{+0.07 \\ -0.09}$ & 1804 & 39 & 232 $^{(a)}$ \\
        NGC\,1365      & 03:34:20.1 & -36:11:07 & 6.4$\substack{+0.06 \\ -0.08}$ & 1691 & 48 & 55 $^{(a)}$  \\
        NGC\,1365      & 03:34:24.9 & -36:07:04 & 7.0$\substack{+0.04 \\ -0.05}$ & 1608 & 34 & 28 $^{(a)}$  \\ 
        NGC\,1365      & 03:34:25.8 & -36:03:46 & 6.8$\substack{+0.05 \\ -0.06}$ & 1597 & 48 & 39 $^{(a)}$  \\
        NGC\,1365      & 03:34:32.3 & -36:15:22 & 6.4$\substack{+0.07 \\ -0.08}$ & 1694 & 59 & 58 $^{(a)}$  \\
        NGC\,1365      & 03:34:34.3 & -36:01:11 & 6.9$\substack{+0.05 \\ -0.06}$ & 1694 & -- & 58 $^{(a)}$  \\
        NGC\,1365      & 03:34:37.1	& -36:25:40 & 6.5$\substack{+0.06 \\ -0.07}$ & 1516 & 36 & 120 $^{a)}$ \\ 
        NGC\,1365      & 03:34:37.3 & -36:10:09 & 6.2$\substack{+0.07 \\ -0.08}$ & 1591 & 32 & 45 $^{(a)}$  \\ 
        NGC\,1365      & 03:34:43.5	& -36:47:30 & 6.3$\substack{+0.08  \\ -0.1}$ & 1591 & -- & 45 $^{(a)}$  \\ 
        NGC\,1365      & 03:34:46.5 & -36:00:39 & 5.7$\substack{+0.11 \\ -0.14}$ & 1591 & -- & 45 $^{(a)}$  \\ 
        NGC\,1365      & 03:34:49.9	& -36:13:50 & 5.8$\substack{+0.09 \\ -0.12}$ & 1636 & 21 & 0 $^{(a)}$   \\
        NGC\,1365      & 03:34:54.6 & -36:49:50 & 6.1$\substack{+0.1  \\ -0.14}$ & 1777 & 28 & 141 $^{(a)}$ \\
        NGC\,1365      & 03:34:55.0 & -36:01:50 & 6.3$\substack{+0.07 \\ -0.09}$ & 1745 & 32 & 109 $^{(a)}$ \\ 
        NGC\,1365      & 03:35:02.3 & -36:22:40 & 6.2$\substack{+0.08 \\ -0.10}$ & 1613 & 62 & 23 $^{(a)}$  \\
        NGC\,1365      & 03:35:03.4 & -36:07:44 & 6.4$\substack{+0.07 \\ -0.08}$ & 1599 & 32 & 37 $^{(a)}$  \\
        NGC\,1365      & 03:35:08.2	& -36:25:27 & 6.2$\substack{+0.07 \\ -0.08}$ & 1633 & 20 & 3 $^{(a)}$   \\   
        NGC\,1365      & 03:35:11.3	& -36:12:12 & 6.3$\substack{+0.07 \\ -0.08}$ & 1569 & 39 & 67 $^{(a)}$  \\
        NGC\,1365      & 03:35:12.2 & -36:31:28 & 6.3$\substack{+0.08 \\ -0.09}$ & 1629 & 30 & 7 $^{(a)}$   \\ 
        NGC\,1365      & 03:35:16.3	& -36:34:14 & 6.8$\substack{+0.05 \\ -0.06}$ & 1623 & 52 & 13 $^{(a)}$  \\
        NGC\,1365      & 03:35:21.0	& -36:32:32 & 6.3$\substack{+0.07 \\ -0.09}$ & 1503 & 43 & 133 $^{(a)}$ \\ 
        NGC\,1365      & 03:35:23.0	& -36:26:55 & 6.4$\substack{+0.07 \\ -0.08}$ & 1596 & 48 & 40 $^{(a)}$  \\ 
        NGC\,1365      & 03:35:22.6	& -36:34:45 & 6.2$\substack{+0.08 \\ -0.11}$ & 1444 & 34 & 192 $^{(a)}$ \\ 
        NGC\,1365      & 03:35:32.7 & -36:18:31 & 6.3$\substack{+0.07 \\ -0.08}$ & 1580 & 27 & 56 $^{(a)}$  \\
        NGC\,1365      & 03:35:37.0	& -36:41:37 & 6.9$\substack{+0.05 \\ -0.06}$ & 1503 & 34 & 133 $^{(a)}$ \\
        NGC\,1365      & 03:36:46.3	& -36:33:49 & 6.3$\substack{+0.08 \\ -0.10}$ & 1563 & 34 & 73 $^{(a)}$  \\
        \hline 
        \hline
    \end{tabular}
    \\
    \textbf{Notes.} (1) Name of nearest galaxy with a measured redshift. (2) RA. (3). Dec. (4) \MHI and its uncertainty as measured from the 41\arcsec-resolution image. (5) The systemic velocity measured in the optical definition. (6) Width of the \HI profile at 20\% of the peak flux. Those without a $w_{20}$ were not able to be determined by the \SoFiA fitting procedure. (7) The velocity difference between the systemic velocities of the clouds and nearest galaxy with a spectroscopic redshift. The $^{(a)}$ denotes a velocity differences calculated from two (cloud and galaxy) MFS detections, and $^{(b)}$ is from a MFS (cloud) and optical (galaxy) $cz$ difference. 

    \label{tab:cloud_detections}
\end{table*}

\end{appendix}

\end{document}